# Astro2020 Science White Paper

# Variability in the Assembly of Protostellar Systems

**Thematic Areas:** ☐ Planetary Systems  ☒ Star and Planet Formation
☐ Formation and Evolution of Compact Objects  ☐ Cosmology and Fundamental Physics
☐ Stars and Stellar Evolution  ☐ Resolved Stellar Populations and their Environments
☐ Galaxy Evolution  ☐ Multi-Messenger Astronomy and Astrophysics

**Principal Authors:**
Name: Joel D. Green
Institution: Space Telescope Science Institute / University of Texas at Austin
Email: jgreen@stsci.edu
Phone: 410-338-4819

Name: Yao-Lun Yang
Institution: University of Texas at Austin
Email: yaolunyang.astro@gmail.com

**Co-authors: (names and institutions)**

Tom Megeath (University of Toledo), Doug Johnstone (National Research Council Canada), John Tobin (National Radio Astronomy Observatory), Sarah Sadavoy (Harvard-Smithsonian Center for Astrophysics), Klaus Pontoppiddan (Space Telescope Science Institute), Stella Offner (The University of Texas at Austin), Neal J. Evans, II (The University of Texas at Austin), Dan M. Watson (University of Rochester), Jennifer Hatchell (University of Exeter), Ian Stephens (Harvard-Smithsonian Center for Astrophysics), Zhi-Yun Li (University of Virginia), Jacob White (Konkoly Observatory), Robert A. Gutermuth (University of Massachusetts Amherst), Will Fischer (Space Telescope Science Institute), Agata Karska (Nicholas Copernicus University), Jens Kauffmann (MIT), Mike Dunham (SUNY Fredonia), Hector Arce (Yale University)

**Abstract:** Understanding the collapse of clouds and the formation of protoplanetary disks is essential to understanding the formation of stars and planets. Infall and accretion, the mass-aggregation processes that occur at envelope and disk scales, drive the dynamical evolution of protostars. While the observations of protostars at different stages constrain their evolutionary tracks, the impact of variability due to accretion bursts on dynamical and chemical evolution of the source is largely unknown. The lasting effects on protostellar envelopes and disks are tracked through multi-wavelength and time domain observational campaigns, requiring deep X-ray, infrared, and radio imaging and spectroscopy, at a sufficient level of spatial detail to distinguish contributions from the various substructures (i.e., envelope from disk from star from outflow). Protostellar models derived from these campaigns will illuminate the initial chemical state of protoplanetary disks during the epoch of giant planet formation. Insight from individual star formation in the Milky Way is also necessary to understand star formation rates in extragalactic sources. This cannot be achieved with ground-based observatories and is not covered by currently approved instrumentation.

**Requirements:** High (v < 10 km/s for survey; v < 1 km/s for followup) spectral resolution capabilities with relatively rapid response times in the IR (3-500 μm), X-ray (0.1-10 keV), and radio (cm) are critical to follow the course of accretion and outflow during an outburst. Complementary, AU-scale radio observations are needed to probe the disk accretion zone, and 10 AU-scale to probe chemical and kinematic structures of the disk-forming regions, and track changes in the dust, ice, and gas within protostellar envelopes.

**Introduction.** Stars and their planetary systems are the end-product of a process that begins with the collapse of a molecular cloud and draws upon the pre-existing mass and energy budget of a 10,000 AU region (see Gutermuth et al. whitepaper). Most of the (low mass; < a few solar masses) protostar's mass accumulates during the first 0.5 Myr (Evans et al. 2009), where protostellar systems are embedded within their birth clouds. Because of dust extinction, peak emission falls in the far-IR for most systems. It is now generally understood that the protostellar evolutionary classes, defined by observations with previous and current generation IR instruments (IRAS, ISO, Spitzer, Herschel, ALMA, SOFIA, JWST, and others), correspond broadly to theoretical stages including a thinning and eventual dissipation of an envelope; formation of a Keplerian disk; accretion of the disk material onto the central star or compaction and collection of solids, or ejection/destruction via radiation pressure; diminishing accretion and outflow processes. These timescales are traced primarily via kinematics, which provides a snapshot in time, and chemistry, which traces the system's evolution (e.g., Kristensen & Dunham 2018). What is the impact of variability on dynamical evolution at the envelope (few thousands AU) scale? How much variability is imposed by instabilities in the physical assembly processes, particularly at disk scales (Dunham et al. 2014; Johnstone 2017; Vorobyov et al. 2018)? Here we focus on the role of time variability and its implications for chemistry and mass accretion.

**What processes regulate accretion, leading to the observed variability in protostars? How will understanding variability change our view of protostellar evolution? What signatures can we monitor with spectroscopy?** Understanding variability in accretion is necessary to interpret protostellar birthline tracks and model the formation of envelopes and disks. We require long-term monitoring of accretion variability from X-ray to radio at high sensitivity, with high resolution IR spectroscopic capabilities, and the potential for daily/weekly/monthly cadence.

**Origins of variability.** The internal evolution of a protostar depends on the manner in which mass is accreted via the protostellar disk and envelope (see Fig 1; Baraffe et al. 2010, 2012; Hosokawa et al. 2011; Kunitomo et al. 2017; Vorobyov et al. 2017; Jensen & Haugbølle 2018). Three main processes of mass and angular momentum transport in disks have been proposed: viscous torques due to turbulence triggered by the magneto-rotational instability (MRI; Balbus and Hawley, 1991) and gravitational torques induced by gravitational instability (GI; Lin & Pringle, 1987; Laughlin & Bodenheimer, 1994), and disk winds (Pudritz et al. 2007). Which mechanism dominates the transport of mass from the disk to the protostar, and which mechanism leads to the observed accretion variability, is currently debated (Audard et al. 2014 and references therein). Variability will alter the density, temperature, and chemistry in the disk, providing a strong diagnostic of the mechanism driving outbursts. Unambiguous measurements of magnetic field strengths and geometry in disks are needed (e.g., Stephens et al. 2018) to understand at which scales angular momentum transfer via MRI and disk winds is effective.

Limited studies of variability in individual systems are beginning to test some of these mechanisms. Recent studies of time evolution in X-ray (e.g. Liebhart et al. 2014), optical (e.g, Lee et al. 2015), IR (e.g, Rebull et al. 2014; Wolk et al. 2015; Furlan et al. 2016; Fischer et al. 2017), and submillimeter spectroscopy (e.g., Logan et al. 2019) are non-systematic, and insufficient to uniquely constrain the regions and timescales and therefore physical origins of variability. Future deeper spectroscopic observations of these sources, using atomic and molecular FIR tracers including [O I] (63 µm) and water (3-500 µm), will provide strong constraints on the lasting effects of outbursts on system chemistry (comparing over decade baselines, e.g., 5-25 µm with Spitzer-SOFIA-JWST; Green et al. 2016; 50-700 µm with Herschel-future FIR; Molyarova et al. 2018).

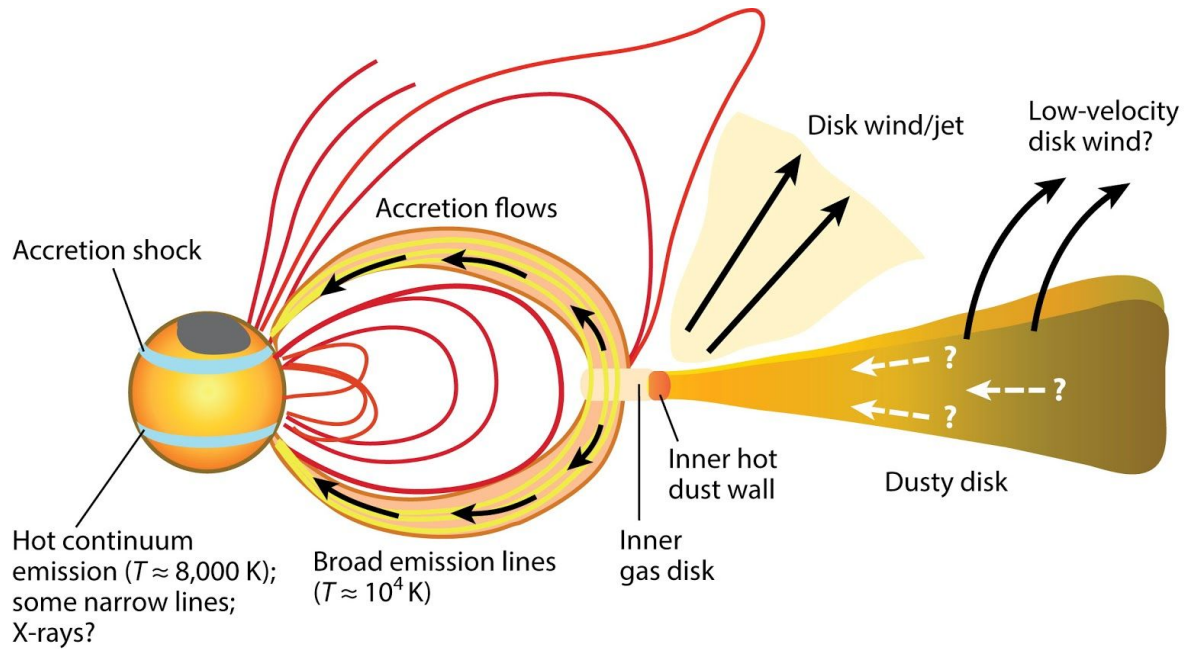

Fig. 1: Disk accretion cartoon, reproduced with permission (Hartmann, Herczeg, & Calvet 2016).

Disk chemistry could be significantly impacted by the accretion history (Fig. 2). First, the location of snow-lines (or ice-lines) in a planet-forming disk is important as an evolving snow-line can greatly impact the abundance of solid material throughout the disk and affect grain growth. ALMA observations of outbursting star V883 Ori show the water snow-line at a radial distance of 42 AU, in contrast with the expected distance of < 5 AU for a solar-like star (Cieza et al. 2016). Therefore, if these eruptive events are a common occurrence in the evolution of protoplanetary disks then their effects on the relocation of snow-lines, and how this can spur grain growth and crystallization suggested by mid-IR studies (e.g., Varga et al. 2018), must be considered in planet formation models. Second, the outburst's impact on chemical pathways must be considered (van't Hoff et al. 2018, Lee et al. 2019). Gas-phase chemistry is one of most efficient way to form complex organic molecules, but in quiescence the time spent in gas-phase is too short to explain the observed high abundances. In outbursts, disks release molecules into the gas phase, such that new and larger complex molecules can be created efficiently, boosting their abundances. Complex organic molecules become optically thick to ALMA in the innermost of regions of disks, where the impact of variability would be largest, but would be optically thin at cm wavelengths; sub-km/s (Ruiz-Rodriguez et al. 2017) resolution at AU-scales at radio wavelengths is needed.

**Need for broad spectroscopic characterization of a range of outbursts.** The viability of dedicated long-term monitoring of this earliest stage of star formation has been proven by several pilot studies, including mid-IR (Fischer, Safron, & Megeath 2019) and sub-mm (JCMT Transients; Herczeg et al. 2017) of eight nearby star-forming regions finds 10% of these low-mass protostars vary in the sub-mm by > 5% per year (Mairs et al. 2017; Johnstone et al. 2018). A new era of multiwavelength photometric variability surveys (e.g., LSST, WFIRST) will address the outburst frequency and duration. But spectroscopy is required to characterize the consequences of outbursts on density, temperature, chemistry, and kinematics. Direct accretion tracers that can track variability of the innermost regions in the youngest embedded sources (e.g. hydrogen lines, CO; Black & Dalgarno 1976; Black & van Dishoeck 1987) await larger ground-based NIR facilities with high spectral resolution (v < few km/s; e.g. Sokal et al. 2018), while FIR to radio spectroscopy will characterize the consequences of outbursts.

Inward motions of gas are the direct signature of the collapse of envelopes. Molecular emission and absorption can selectively trace ranges of density and temperature throughout the envelope. Therefore, wide spectral coverage is essential to take advantage of the full molecular inventory to measure the dynamics and chemistry in reactions to the accretion variability. In radio, high spatial resolution and wide bandwidth facilities will allow us to probe complex molecules from the densest part of disk, where the high optical depth hinders such observation at shorter wavelengths. Simple molecules like $C^{18}O$, $^{13}CO$, $HCO^+$, and HCN trace the envelope kinematics; widened submm bandwidth provides a suite of complex molecules that trace several specific regions in a large sample, once we identify these molecules unambiguously with supported laboratory experiments (e.g., Willacy, Allen, & Yung 2016). One issue may be optical depth, which requires longer wavelengths. Thus capabilities at cm wavelengths will be important for tracing inner envelopes in the youngest sources.

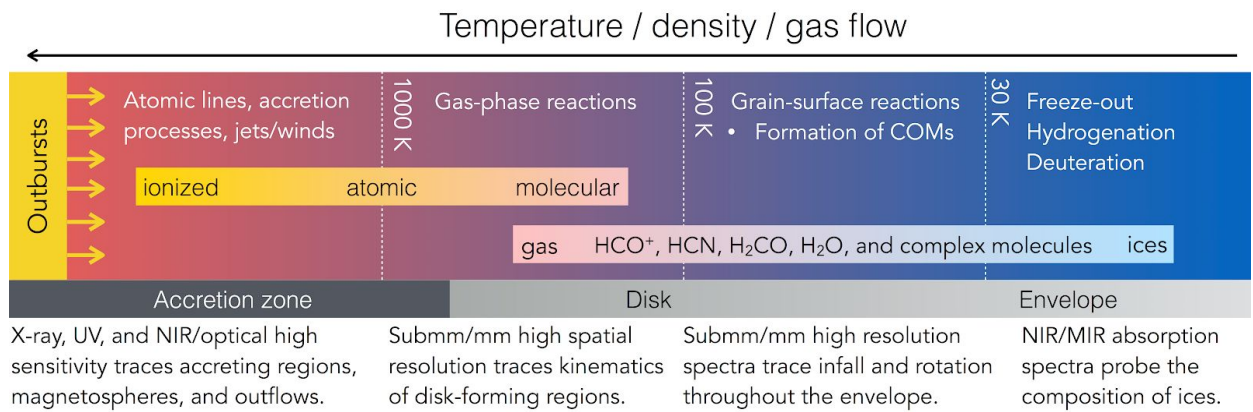

Fig. 2: Representative chemical and kinematic zones in protostars (courtesy: Y.-L. Yang).

The current chemical state we observe – now detecting numerous complex organic molecules (e.g., Jørgensen et al., 2018; Yang et al., in prep.) -- changes through interactions, highlighting the extensive chemical evolution during the first 0.5 Myr (e.g., Jørgensen et al., 2016). In the cold and dense regions of cores, gas freezes onto dust grains as ices, which enable the formation of larger molecules via gas-grain chemistry (e.g., Garrod et al. 2017; Aikawa et al. 2018). Ices also make dust grains grow more efficiently. As the temperature rises due to infall or accretion, different chemical reactions occur on the dust grains, resulting in a variety of chemical landscapes when those ices are thermally desorbed. Multiple points in the chemical history can be probed by different observatories. JWST and SPHEREx, for example, can probe pristine ices, and ALMA can follow the gas-phase evolution in the disk, but the journey in between, from 10 to 100 K, is largely unknown. FIR spectroscopy will best observe molecules in this temperature range, particularly water. This will require unbiased surveys over relatively short observing times to understand environmental and evolutionary effects.

Spatial resolution becomes a critical factor in understanding the real-time effects of eruptive events. This will require high spectral or spatial resolution at moderate sensitivity across a variety of wavelengths, particularly the far-IR/mm/radio wavelengths as well as high energy for outflow/shock/accretion processes (e.g., Manoj et al. 2016), and the maintenance of current capabilities at these wavelengths into the future. For example, a resolution of 0.7 mas corresponds to spatial scales of 0.25 and 0.10 AU for two prototypical outburst systems, FU Ori and EX Lup, respectively, roughly coincident with their accretion zones (Selina et al. 2018), proposed for the longest baselines of the ngVLA (see White et al. whitepaper).

**Need for X-ray diagnostics of accretion and outflow.** Accretion variability should also be reflected in jet variability (Konigl & Pudritz 2000) as the disk or X-wind responds to configuration changes and varying ionization levels. Jet rotation is a fundamental prediction of magnetically-driven wind models, in which angular momentum is carried away by jets. Higher spatial and spectral resolution on jets will measure rotation, as suggested by Bacciotti et al. (2002) for young star DG Tau using HST. Arce et al. (2007) argued that the observed sequences of knots in jets provides a fossil record of the episodic accretion onto the protostar, a result followed up by theoretical investigation (Vorobyov et al. 2018; Rohde et al. 2019). X-ray studies of protostars (e.g., Guedel et al. 2007, 2008; Skinner et al. 2010) hint at multi-component emission: coronal, magnetospheric, jet-driven activity, and stellar rotation (Skinner, Audard, & Guedel 2016). Larger collecting areas and higher spectral resolution will enable disambiguation of models of blended highly ionized line emission that diverge fundamentally on the state of the emitting gas in the inner/coronal regions near the origin of the launched material.

**Need for multi-wavelength simultaneity.** Systems with multiple interaction scales invariably require multi-wavelength studies to reveal the underlying physics. For example, simultaneous radio and X-ray observations in X-ray binaries (e.g., Corbel et al. 2003) have enabled understanding of fundamental physics of mass scaling in black holes (Falcke, Koerding, & Markoff 2004). YSO outburst rising times can be weeks to months, and radio flares (e.g., Mairs et al. 2019) are likely accompanied by enhanced UV/X-ray emission. Similarly, multi-wavelength monitoring during a burst can track the progress of the temperature rise from the inside out, through the disk and envelope via mid- and far-IR (peak energy near 100 μm; Green et al. 2013). This temperature rise is reflected in the dust emission (> a few μm), propagating to longer wavelengths as the extent of heating increases (Johnstone et al. 2013).

**Need for high spectral and spatial resolution observations of the youngest protostars.** The rate of identification of outbursting objects has accelerated in the past decade, due to increased coverage. With anticipated survey capability out to 1.5 kpc, assuming protostars spend 0.1-1% of their time in outburst, we will likely flag about 1-10 outbursting young stars per year within 1.5 kpc in the 2030s; thus we need to understand in detail how different components react to the outburst with panchromatic spectroscopic observations. Variability undoubtedly occurs in earlier phases of star formation than currently observed (mostly Class

I/II), but the optical to mid-IR properties of Class 0 sources are very poorly sampled with current generation of instruments: space-based infrared telescopes like Spitzer had wide spectral coverage but with low spectral resolution instrumentation; ground-based telescopes have high resolution spectroscopic capabilities but lack the sensitivity to detect the optical to mid-IR emission that originates from the innermost regions of protostellar systems, including common tracers of magnetospheric accretion that would indicate whether accretion in very young systems occurs in the same fashion as more evolved ones (Hartmann, Herczeg, & Calvet 2016). The IR flux density of outbursting sources within 1 kpc can peak in excess of 10 Jy; thus high saturation limits will expand our ability to respond rapidly to all outburst detections.

Spitzer and other observatories probed the ices at the core scale and ALMA measures the gas-phase complex molecules in the inner envelopes and disks, but the transition from ices to complex molecules is largely unknown. Where observed, the spatial and spectral resolution is very coarse at wavelengths observed from space, making it challenging to separate the contributions of the envelope and disk substructures. A large scale low resolution infrared survey could focus on ice features in perhaps hundreds of sources including those at distances greater than 1 kpc, which do not require high spectral resolution (R = few hundred is adequate), in an unbiased, all-sky fashion. At high resolution, a combination of IR to mm surveys combined with high fidelity followup (i.e., JWST and future observatories) on tens of protostars would provide benchmarks for the chemical evolutionary process and statistically dynamical evolution. An ALMA upgrade to wider bands will greatly increase the efficiency of chemical surveys. Observatories like ALMA are revealing complex organic molecules that new chemical models-based lab astrophysics data will be needed to interpret. These studies on a handful of sources suggest species and trends for follow-up with telescopes that will have large increases in sensitivity at high spectral resolution over many more targets in the IR, or achieve larger surveys at submm/mm wavelengths.

**Need for improved atomic/molecular models.** A key prerequisite to the chemical evolution models is continued progress in atomic and molecular collisional-excitation rate coefficients. The ones that are accurate and reliable (e.g., CO, [O I]) produce degenerate models; the ones which could in principle break these degeneracies (e.g., $H_2O$, [Fe II]) have never been subjected to constraints from laboratory measurements and have been changing rapidly in the literature (e.g., Ramsbottom et al. 2007; Bautista et al. 2015). The situation is even more challenging for complex organic molecules.

**Conclusion.** High (v < 10 km/s for survey; v < 1 km/s for followup) spectral resolution capabilities with relatively rapid response times in the IR (3-500 µm), X-ray (0.1-10 keV), and radio (cm) are critical to follow the course of accretion and outflow during an outburst. Complementary, AU-scale radio observations are needed to probe the disk accretion zone, and 10 AU-scale to probe chemical and kinematic structures of the disk-forming regions, and track changes in the dust, ice, and gas within protostellar envelopes. This cannot be achieved with ground-based observatories and is not covered by currently approved instrumentation.